\documentclass[%
 reprint,
 amsmath,amssymb,
 aps,
showkeys
]{revtex4-1}

\usepackage{graphicx}
\usepackage{dcolumn}
\usepackage{bm}

\usepackage{xcolor}

\begin{document}


\title{Revisiting and revising Tatarskii's formulation for the temperature structure parameter ($C_T^2$) in atmospheric flows}

\author{Sukanta Basu}
\email{sukanta.basu@gmail.com}
\affiliation{Faculty of Civil Engineering and Geosciences, Delft University of Technology, Delft, the Netherlands
}
\author{Albert A. M. Holtslag}
\email{bert.holtslag@wur.nl}
\affiliation{Meteorology and Air Quality, Wageningen University, Wageningen, the Netherlands}%

\date{\today}

\begin{abstract}
In this paper, we revisit a well-known formulation of temperature structure parameter ($C_T^2$), originally proposed by V.~I.~Tatarskii. We point out its limitations and propose a revised formulation based on turbulence variance and flux budget equations. Our formulation includes a novel physically-based outer length scale which can be estimated from routine meteorological data.
\end{abstract}


\maketitle

\section{Introduction}
\label{intro}

According to the hypothesis by Kolmogorov-Obukhov-Corrsin \citep[KOC;][]{corrsin51,kolmo41,obukhov49} within the inertial-range, the second-order structure function ($S_{2\theta}^{KOC}$) of potential temperature ($\theta$) in the vertical direction ($z$) should behave as follows:
\begin{equation}
\label{KOC1}
S_{2\theta}^{KOC} = \langle \left(\Delta \theta\right)^2\rangle  \sim  \left(\overline{\varepsilon}\right)^{-1/3} \overline{\chi}_{\theta} \left(\Delta z\right)^{2/3}. 
\end{equation}
Here $\Delta z$ is a separation distance that varies within the inertial range $l_0 \ll \Delta z \ll \Lambda_0$. The inner and outer scales of turbulence are denoted by $l_0$ and $\Lambda_0$, respectively. The angular bracket and overlines denote ensemble averaging. The dissipation rates of turbulent kinetic energy (TKE; $\overline{e}$) and variance of temperature ($\sigma_\theta^2$) are denoted by $\overline{\varepsilon}$ and $\overline{\chi}_{\theta}$, respectively. 

In the literature, Eq.~(\ref{KOC1}) is commonly re-written as follows: 
\begin{subequations}
\begin{equation}
S_{2\theta}^{KOC} = C_T^2 \left(\Delta z\right)^{2/3},  
\label{KOC2}
\end{equation}
where $C_T^2$ is the so-called temperature structure parameter which is commonly expressed as \cite{wyngaard71}: 
\begin{equation}
C_T^2 = \left(\frac{c}{2} \right)~\left(\overline{\varepsilon}\right)^{-1/3} \overline{\chi}_{\theta} = c \left(\overline{\varepsilon}\right)^{-1/3} \overline{N}_{\theta}.
\label{CT2a}
\end{equation}
\end{subequations}
The variable $\overline{N}_{\theta}$ is simply one half of $\overline{\chi}_{\theta}$ and has been used instead of $\overline{\chi}_{\theta}$ in a number of publications (including by Tatarskii~\cite{tatarskii71}). The proportionality constant $c$ is typically assumed to be in the range of 2.8--3.2. 

Direct measurement of $\overline{\varepsilon}$, $\overline{\chi}_{\theta}$ (or, $\overline{N}_{\theta}$), and $C_T^2$ is difficult in the atmosphere as it involves computations of fine-scale velocity and temperature gradients with spatial resolutions of a few mm (i.e., on the order of $l_0$). As a viable alternative, fifty years ago, Tatarskii~\cite{tatarskii71} formulated the following equation: 
\begin{equation}
    C_T^2 = \left(\frac{c}{Pr_t}\right) L_0^{4/3} \Gamma^2, 
    \label{CT2b}
\end{equation}
where $\Gamma$ is the gradient of mean potential temperature (i.e., $\partial\overline{\theta}/\partial z$). $L_0$ is a characteristic length scale; its relationship with $\Lambda_0$ will be discussed later. $Pr_t$ is known as the turbulent Prandtl number. For simplicity, several studies \cite[e.g.,][]{beland88,vanzandt78,walters81} in the past have (incorrectly) assumed $Pr_t$ to be equal to unity.

According to Eq.~(\ref{CT2b}), if $L_0$ and $\Gamma$ can be reliably measured, estimated or prescribed, then $C_T^2$ values can be computed directly. Since it is relatively straightforward to measure or estimate $\Gamma$ in the atmosphere, over the past decades, several studies focused on the estimation of $L_0$. Some of these studies  \cite[e.g.,][]{coulman88,dewan93,vanzandt78} proposed empirical regression equations. Others \cite{basu15,wu20} recommended usage of certain well-known length scales (e.g., Thorpe scale, Ellison scale) as surrogates of $L_0$.   

The purpose of the present study is two-fold. First, we revisit the derivation of Eq.~(\ref{CT2b}) following the footsteps of Tatarskii~\cite{tatarskii71}. Unfortunately, his original derivation was somewhat cryptic and included various implicit assumptions. We have attempted to elucidate on the derivation and the assumptions to the best of our abilities. In addition, we have made a sincere effort to point out some fundamental limitations of Tatarskii's assumptions in the context of atmospheric flows. Next, by relaxing some of these assumptions, we have derived an alternative to Eq.~(\ref{CT2b}) from fundamental budget equations of stably stratified flows. In our formulation, $L_0$ is directly related to common meteorological variables (e.g., variance of potential temperature) and does not require any ad-hoc prescription. 

The structure of the paper is as follows. In Section~\ref{Tata}, we delve into the original derivation of Eq.~(\ref{CT2b}). Its association with turbulence scaling laws in the buoyancy-range is explored in Section~\ref{Buoyancy}. Our proposed formulation is documented in Section~\ref{BH21}. The limitations of both Tatarskii's and our proposed formulations for convective mixed layer are discussed in Section~\ref{CBL}. The findings of this study are summarized in Section~\ref{Conc}. Finally, in Appendix~1, we derive formal relationships among certain length scales. 

\section{Derivation of Tatarskii's $C_T^2$ Formulation}
\label{Tata}

Tatarskii utilized both heuristic arguments and turbulence variance budget equations for the derivation of Eq.~(\ref{CT2b}). They are documented in sections 14 and 15 of \cite{tatarskii71}. At the outset, it is important to note that these derivations do not account for the buoyancy effects (discussed below). However, in section 17 of \cite{tatarskii71}, buoyancy effects are mentioned in a different context; there, Tatarskii derived a diagnostic relationship between $L_0$ and $\Lambda_0$ following a formalism by Bolgiano~\cite{bolgiano59,bolgiano62} and Obukhov~\cite{obukhov59}. In this paper, we elaborate on all these derivations albeit with additional clarifications. Furthermore, we have slightly changed some of the notations used by Tatarskii to be consistent with the contemporary literature and also with the rest of our manuscript. For better logical reasoning, we have also changed the sequential ordering of equations used by Tatarskii.

\subsection{Heuristic Arguments}

 The following text has been taken verbatim from Section 14 (page 73) of \cite{tatarskii71}:  

\begin{quotation}
``A gradient of mean temperatures will result in a systematic temperature difference between any two points at different heights. This temperature difference $\Delta \theta$ is approximately given by $\Delta \theta \approx \Gamma \Delta z$, and its square is $\left(\Delta \theta\right)^2 \approx \Gamma^2 \left(\Delta z\right)^2$. There is also a random temperature difference between these points, whose mean square value is $C_T^2 \left(\Delta z \right)^{2/3}$. For small $\Delta z$, $C_T^2 \left(\Delta z \right)^{2/3}$ is much greater than the gradient term $\Gamma^2 \left(\Delta z\right)^2$ (i.e., the random temperature differences are much greater than the systematic ones). There is however a certain $\Delta z_0$ when the two factors become comparable, and for $\Delta z > \Delta z_0$ the mean temperature difference is greater than the random difference. This $\Delta z_0$ is interpreted as the vertical mixing scale. Clearly the ``2/3 law'' is applicable only over distances $\Delta z$ not greater than this mixing scale. Therefore $\Delta z_0$ may be taken equal to the outer scale of turbulence.''
\end{quotation}
Essentially, Tatarskii suggested equating the inertial-range term (`random') with a larger-scale term (`systematic') as follows:
\begin{subequations}
\begin{equation}
    C_T^2 \left(\Delta z_0 \right)^{2/3} = \Gamma^2 \left(\Delta z_0\right)^2,
    \label{Heur1}
\end{equation}
\begin{equation}
    \mbox{or, } C_T^2 =  \left(\Delta z_0\right)^{4/3} \Gamma^2.
    \label{Heur2}
\end{equation}
\end{subequations}
It is needless to say that Eq.~(\ref{Heur2}) is virtually identical to Eq.~(\ref{CT2b}) as long as $\Delta z_0$ is proportional to $L_0$. In a latter section of this paper, we will show that an equation similar to Eq.~(\ref{Heur2}) can also be derived by making use of the vertical scaling characteristics of temperature and wind speed in the buoyancy-range ($\Delta z \gg \Lambda_0 $).

Even though the heuristic arguments can in general provide valuable insights, in turbulence research, more rigorous results can be obtained via variance and flux budget equations. Tatarskii \cite{tatarskii71} made great strides in this direction as discussed next. 

\subsection{Turbulence Variance Budget Equations}

Tatarskii's derivation for Eq.~(\ref{CT2b}) is for a specific type of turbulent flow: uniform shear flow  with an imposed temperature gradient. He further (implicitly) considered temperature to be a passive scalar. This assumption implies that the dynamical evolution of temperature is driven by the velocity field; however, temperature does not modulate the velocity field in any manner. A direct consequence of this assumption is that the buoyancy terms are neglected in the Navier-Stokes momentum and TKE equations. 

In line with non-buoyant, shear flow turbulence literature, Tatarskii assumed the inverse of velocity shear to be a characteristic time-scale; where, velocity shear ($S$) is: 
\begin{equation}
    S = \left(\frac{\partial \overline{u}}{\partial z}\right).
\end{equation}
Here, $\overline{u}$ is the velocity component along the mean wind direction. If $L_0$ is the characteristic length scale, then the velocity scale is simply: $L_0 S$. Then, the eddy viscosity coefficient ($K_M$), commonly parameterized as a product of characteristic length and velocity scales, can be expressed as: 
\begin{equation}
    K_M = L_0 \left(L_0 S\right) = L_0^2 S. 
    \label{KM}
\end{equation}
This is Eq.~(14.23) of \cite{tatarskii71}.

Under the assumption of steady-state and horizontal homogeneity, the simplified budget equation for TKE ($\overline{e}$) can be written as:
\begin{equation}
    \overline{\varepsilon} = - \left( \overline{u' w'} \right) S. 
    \label{EDR1}
\end{equation}
Here, the left hand and right hand sides of this equation represent (molecular) dissipation and shear production of turbulence, respectively. The along-wind component of the momentum flux is $\overline{u' w'}$. As mentioned earlier, owing to the passive scalar assumption, the buoyancy term is not included in Eq.~(\ref{EDR1}). Furthermore, the terms with secondary importance (e.g., turbulent transport) are neglected.  

According to the K-theory, based on the celebrated hypothesis of Boussinesq in 1877, the momentum flux can be approximated as follows: 
\begin{equation}
    \overline{u' w'} = - K_M S. 
    \label{uw}
\end{equation}
By combining Eqs.~(\ref{EDR1}) and (\ref{uw}), we get: 
\begin{equation}
    \overline{\varepsilon} = K_M S^2. 
    \label{EDR2}
\end{equation}
This is Eq.~(14.21) of \cite{tatarskii71}.

Similar to Eq.~(\ref{EDR1}), the simplified budget equation of potential temperature variance ($\sigma_\theta^2$) can be written as follows: 
\begin{equation}
    \overline{N}_{\theta} = - \left( \overline{w' \theta'} \right) \Gamma, 
    \label{CHI1}
\end{equation}
where $\overline{w' \theta'}$ denotes the sensible heat flux. Once again, by using the K-theory, we can write: 
\begin{equation}
    \overline{w' \theta'} = - K_H \Gamma. 
    \label{wT}
\end{equation}
Here $K_H$ is the eddy-diffusivity coefficient and the ratio of $K_M$ and $K_H$ is called the turbulent Prandtl number ($Pr_t$). By combining Eqs.~(\ref{CHI1}) and (\ref{wT}), we get:
\begin{equation}
    \overline{N}_{\theta} = K_H \Gamma^2 = \left(\frac{K_M}{Pr_t}\right) \Gamma^2.
    \label{CHI2}
\end{equation}
This is Eq.~(14.20) of \cite{tatarskii71}.

Now, via Eqs.~(\ref{CT2a}), (\ref{KM}), (\ref{EDR2}), and (\ref{CHI2}), we can deduce Eq.~(\ref{CT2b}) as follows: 
\begin{widetext}
\begin{equation}
C_T^2 = c \left( K_M S^2 \right)^{-1/3} \left(\frac{K_M}{Pr_t}\right) \Gamma^2 = \left(\frac{c}{Pr_t} \right) \left( \frac{K_M}{S} \right)^{2/3} \Gamma^2 = \left(\frac{c}{Pr_t} \right) L_0^{4/3} \Gamma^2. 
\end{equation}
\end{widetext}
These equations are documented as Eqs.~(14.22) and (14.24) in \cite{tatarskii71}.

It is well accepted in turbulence literature that both shear and buoyancy deform the larger eddies more compared to the smaller ones \cite[e.g.,][]{smyth00,basu21a}. Corrsin \cite{corrsin58} postulated that the eddies smaller than $L_C = \left( \frac{\overline{\varepsilon}}{S^3} \right)^{1/2}$ are not affected by shear.  
It is important to recognize that Tatarskii's length scale $L_0$ is actually Corrsin's length scale $L_C$. Based on Eqs.~(\ref{KM}) and (\ref{EDR2}), we can show that:
\begin{equation}
    L_0 = \left( \frac{K_M}{S} \right)^{1/2} = \left( \frac{\overline{\varepsilon}}{S^3} \right)^{1/2} = L_C. 
    \label{LC}
\end{equation}
In shear flow turbulence, the eddies can be assumed to be isotropic if they are smaller than $L_C$ \cite{corrsin58,basu21a}. However, for buoyancy-dominated flows, $L_C$ (and $L_0$) may not be a relevant length scale. Tatarskii realized this possibility and advocated for a specific buoyancy length scale in Section 17 of \cite{tatarskii71}. In the following section, we discuss Tatarskii's strategy and point out its limitations.   

\section{Buoyancy-Range Scaling}
\label{Buoyancy}

\subsection{Hypotheses of Bolgiano and Obukhov}

\begin{figure*}[ht]
\includegraphics[height=1.9 in]{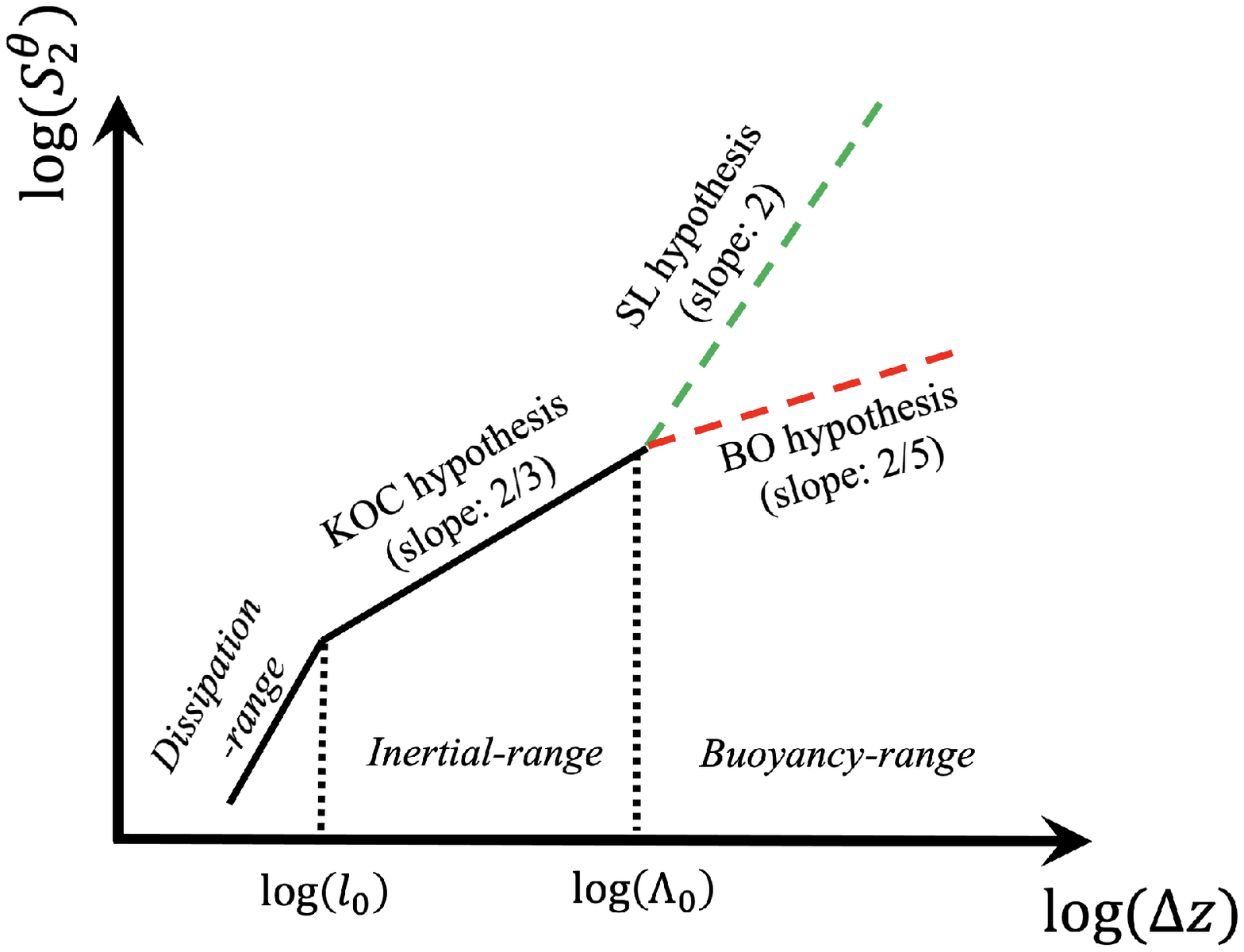}
\includegraphics[height=1.9 in]{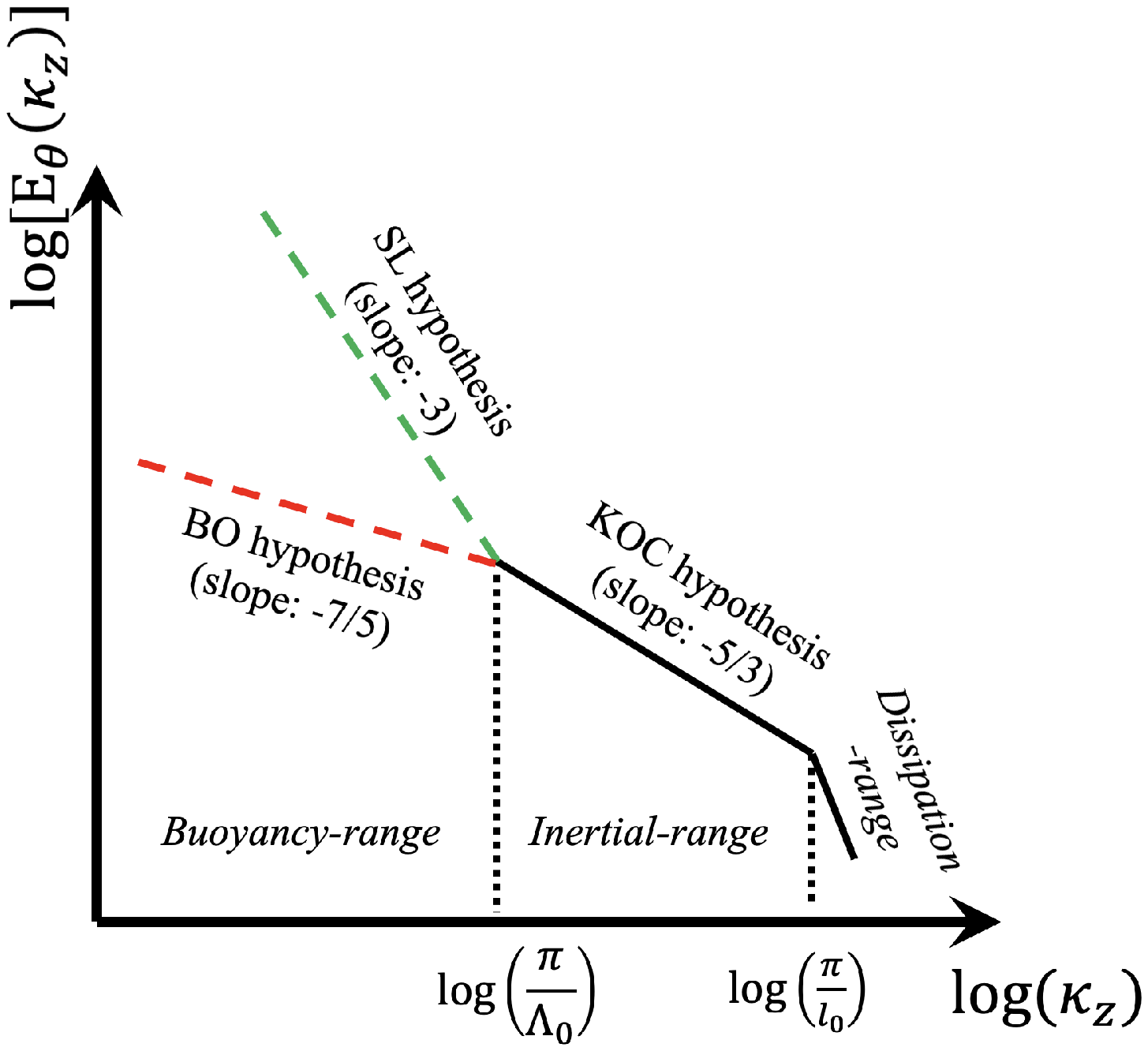}
\caption{Scaling regimes of atmospheric turbulence. The left and right panels represent second-order structure function and spectra of potential temperature, respectively. The scaling hypothesis by Kolmogorov-Obukhov-Corrsin (KOC) holds in the inertial-range. However, various competing hypotheses exist for the buoyancy range. The scaling relationships by Bolgiano-Obukhov (BO) and Shur-Lumley (SL) are compared in these schematics.}
\label{schematic}
\end{figure*}

Over the years, various scaling-based hypotheses have been proposed for the buoyancy-range (scales larger than the inertial-range) of turbulence. One popular hypothesis was independently formulated by Bolgiano~\cite{bolgiano59,bolgiano62} and Obukhov~\cite{obukhov59}. They (henceforth Bolgiano-Obukhov or BO) postulated that in the buoyancy range (i.e., $\Delta z \gg \Lambda_0$), the relevant variables are $\overline{N}_\theta$, and $\beta$; where the buoyancy parameter ($\beta$) is defined as follows: 
\begin{equation}
\beta = \left(\frac{g}{\Theta_0}\right).
\end{equation}
Here the gravitational acceleration is denoted by $g$ and $\Theta_0$ is a reference potential temperature. Via simple dimensional analysis, BO derived the following relationship for second-order structure function of potential temperature in the buoyancy range: 
\begin{equation}
S_{2\theta}^{BO} \sim \left(\overline{N}_\theta\right)^{4/5} \left( \beta \right)^{-2/5} \left( \Delta z \right)^{2/5}.
\label{BO1}
\end{equation}
A schematic of this relationship is shown in the left panel of Fig.~\ref{schematic}. In the right panel of this figure, the BO formulation for potential temperature spectra is depicted.   

From the schematics in Fig.~\ref{schematic}, it is evident that at the cross-over point between the inertial-range and the buoyancy-range (i.e., $\Delta z = \Lambda_0^{BO}$), both Eqs.~(\ref{KOC1}) and (\ref{BO1}) should hold. As a result, we have: 
\begin{subequations}
\begin{equation}
\left(\overline{\varepsilon}\right)^{-1/3} \overline{N}_{\theta} \left(\Lambda_0^{BO}\right)^{2/3} \sim \left(\overline{N}_\theta\right)^{4/5} \left( \beta \right)^{-2/5} \left( \Lambda_0^{BO} \right)^{2/5}.    
\label{BO2}
\end{equation}
By algebraic manipulation, we get a length scale: 
\begin{equation}
  \Lambda_0^{BO} \equiv \left(\overline{\varepsilon}\right)^{5/4} \left(\overline{N}_\theta\right)^{-3/4} \left( \beta \right)^{-3/2}.
  \label{BO3}
\end{equation}
In the literature, this specific length scale is known as the Bolgiano-Obukhov length scale. Tatarskii referred to this (buoyancy) length scale in Eq.~(17.1) of \cite{tatarskii71}. 
\end{subequations}

Tatarskii \cite{tatarskii71} recognized that $L_0$ is essentially a free parameter in his formulation [i.e., Eq.~(\ref{CT2b})] and should be prescribed or estimated separately. In order to circumvent this limitation, Tatarskii derived a formal relationship between $L_0$ and a more physically-based length scale $\Lambda_0^{BO}$ as discussed below. Since Eq.~(\ref{EDR1}) does not include the buoyancy term, as a first step, Tatarskii used a revised budget equation: 
\begin{subequations}
\begin{equation}
    \overline{\varepsilon} = - \left( \overline{u' w'} \right) S + \left(\overline{w'\theta'}\right) \beta. 
    \label{EDR3}
\end{equation}
The second term on the right hand side signifies production (unstable condition) or dissipation (stably stratified condition) of TKE due to the buoyancy effects. Using the K-theory [refer to Eqs.~(\ref{uw}) and (\ref{wT})] and the definition of turbulent Prandtl number ($Pr_t$), this equation can be rewritten as follows: 
\begin{equation}
    \overline{\varepsilon} = K_M \left(S^2 - \frac{\Gamma \beta}{Pr_t} \right) = K_M S^2 \left(1 - \frac{Ri_g}{Pr_t} \right),
    \label{EDR4}
\end{equation}
where $Ri_g$ $\left(= \beta \Gamma/S^2 \right)$ is the gradient Richardson number. It is a non-dimensional variable which quantifies the strength of atmospheric stability. Non-buoyant conditions are characterized by $Ri_g = 0$.
\end{subequations}

Now, by utilizing Eqs.~(\ref{CHI2}), (\ref{LC}), (\ref{BO3}), and (\ref{EDR4}), the length scale $\Lambda_0^{BO}$ can be re-written as: 
\begin{widetext}
\begin{subequations}
\begin{equation}
    \Lambda_0^{BO} = \left( \frac{K_M}{S} \right)^{1/2} \left[ \frac{Pr_t^{3/4} \left(1 - Ri_g/Pr_t \right)^{5/4}}{Ri_g^{3/2}} \right] = L_0 \left[ \frac{Pr_t^{3/4} \left(1 - Ri_g/Pr_t \right)^{5/4}}{Ri_g^{3/2}} \right],
    \label{BO4}
\end{equation}
or,
\begin{equation}
L_0 = \Lambda_0^{BO} \left[ \frac{Ri_g^{3/2}}{Pr_t^{3/4} \left(1 - Ri_g/Pr_t \right)^{5/4}} \right]. 
\label{BO5}
\end{equation}
\end{subequations}
\end{widetext}
This is essentially Eq.~(17.10) of \cite{tatarskii71}. 

In principle, if one can measure $\Lambda_0^{BO}$, then $L_0$ can be estimated via Eq.~(\ref{BO5}). However, direct estimation of $\Lambda_0^{BO}$, in our atmosphere, is challenging (if not impossible) since it requires reliable estimates of both $\overline{\varepsilon}$ and $\overline{N}_\theta$. Apart from this practical limitation, there are more fundamental issues. 

In the literature, only a handful of laboratory and numerical studies of idealized flows \cite[e.g.,][]{boffetta12,kumar14,niemela00,verma18} have ever documented the existence of the BO scaling. However,  to the best of our knowledge, the observational support for this scaling hypothesis in our atmosphere is non-existent. Thus, the usage of $\Lambda_0^{BO}$ as an outer (cross-over) scale is questionable.  

Another critical issue is the incompatibility of $\Lambda_0^{BO}$ with Eq.~(\ref{CT2b}). Using Eq.~(\ref{CT2a}), it is straightforward to manipulate Eq.~(\ref{BO3}) to arrive at: 
\begin{equation}
    C_T^2 = \left( \overline{N}_\theta \right)^{4/5} \left(\beta \right)^{-2/5} \left( \Lambda_0^{BO}\right)^{-4/15}.
\end{equation}
It is needless to say that this equation is distinctively different from Eq.~(\ref{CT2b}). In other words, Tatarskii's strategy to couple the BO hypothesis with his own $C_T^2$ formulation is not justified. However, we can make use of a different scaling formalism to resolve this inconsistency as elaborated below. 

\subsection{Hypotheses of Shur, Lumley, Monin, and Weinstock}

In the past, several studies \cite[e.g.,][]{cot89,hovde11,nastrom97,tsuda91} have performed 1-D spectral and structure function analyses of vertical (potential) temperature  and wind speed profiles measured by radiosondes and dropsondes in the atmosphere. They often reported a quasi-universal $\kappa_z^{-3}$ spectral scaling behavior over extended vertical ranges; where $\kappa_z$ is the vertical wavenumber. Similarly, in the physical space, the second-order structure function was found to scale as $\left(\Delta z\right)^2$. Please refer to the schematics in Fig.~\ref{schematic} for a summary of various scaling regimes and associated hypotheses. 

To explain the observed vertical scaling behaviors of (potential) temperature and wind speed profiles, several competing hypotheses have been proposed in the literature \cite[e.g.,][]{dewan97,lumley64,monin65,shur62,weinstock85}. In this paper, we discuss the hypothesis of Shur~\cite{shur62} and Lumley~\cite{lumley64} (henceforth the SL hypothesis) because of its simplicity. Monin~\cite{monin65} further simplified the SL hypothesis by using dimensional analysis and we follow his approach. He assumed that, in the buoyancy-range of turbulence, $\beta$ and $\Gamma$ are the only relevant variables. Then, via dimensional analysis, it is trivial to show that the second-order structure functions should behave as follows: 
\begin{subequations}
\begin{equation}
    S_{2u}^{SL} \sim \beta \Gamma \left(\Delta z\right)^2,
    \label{SLu}
\end{equation}
\begin{equation}
    S_{2\theta}^{SL} \sim \beta^0 \Gamma^2 \left(\Delta z\right)^2 = \Gamma^2 \left(\Delta z\right)^2.
    \label{SLth}
\end{equation}
\end{subequations}
According to Kolmogorov's hypothesis \cite{kolmo41}, the following relationship holds in the inertial-range: 
\begin{equation}
    S_{2u}^{K41} \sim \varepsilon^{2/3} \left(\Delta z\right)^{2/3}.
    \label{K41u}
\end{equation}

By matching Eq.~(\ref{SLu}) with Eq.~(\ref{K41u}) at the cross-over point (i.e., $\Delta z = \Lambda_0^{SL}$), we get:
\begin{subequations}
\begin{align}
    \beta \Gamma \left(\Lambda_0^{SL} \right)^2 & \equiv \varepsilon^{2/3} \left(\Lambda_0^{SL}\right)^{2/3},\\
\mbox{or, \hspace{0.1in}}  \left(\Lambda_0^{SL} \right)^{4/3} & \equiv \left(\frac{\varepsilon^{2/3}}{\beta \Gamma} \right) = \left(\frac{\varepsilon^{2/3}}{N_{BV}^2} \right),\\
\mbox{Therefore, \hspace{0.1in}} \Lambda_0^{SL} & \equiv \left( \frac{\varepsilon}{N_{BV}^3}\right)^{1/2} = L_{OZ}.
\label{LOZ}
\end{align}
\end{subequations}
Here $N_{BV}$ is the so-called Brunt-V\"ais\"al\"a frequency. The length scale $L_{OZ}$ was first proposed by Ozmidov~\cite{ozmidov1965turbulent}. 

In a similar manner, by matching Eq.~(\ref{SLth}) against Eq.~(\ref{KOC2}) at $\Lambda_0^{SL}$, one can deduce: 
\begin{subequations}
\begin{equation}
    C_T^2 \left(\Lambda_0^{SL} \right)^{2/3} \equiv \Gamma^2 \left(\Lambda_0^{SL}\right)^2.
\end{equation}
Thus, 
\begin{equation}
 C_T^2 \equiv \left(\Lambda_0^{SL}\right)^{4/3} \Gamma^2 = \left(L_{OZ} \right)^{4/3} \Gamma^2.
\end{equation}
\end{subequations}
This scaling-based equation is almost identical to Eq.~(\ref{CT2b}) if $L_0$ is proportional to $L_{OZ}$ and $Pr_t$ is on the order of one. 

In summary, the Bolgiano-Obukhov scaling is not in agreement with Tatarskii's $C_T^2$ formulation. Instead, in a non-rigorous manner, Tatarskii's formulation can be deduced from the well-known vertical scaling characteristics of (potential) temperature  and wind speed profiles. In the following section, we will take a completely different route. Based on the rigorous turbulence variance and flux budget equations, we will derive a revised $C_T^2$ formulation along with a physically-based length scale formulation. We want to emphasize that in contrast to Tatarskii's approach, our formulation does not require an ad-hoc prescription of an outer length scale; the length scale is an integral part of our analytical derivations.   

\vspace{0.2in}
\section{Revised $C_T^2$ Formulation}
\label{BH21}

In a recent study \cite{basu21c}, we have derived closed-form solutions for a characteristic length scale (denoted as $L_X$) and turbulent Prandtl number ($Pr_t$) based on the conventional budget equations. For brevity, we only report the key findings from \cite{basu21c} which are directly relevant for the present paper. In \cite{basu21c}, we used Eqs.~(\ref{CHI1}), (\ref{EDR4}), and an additional budget equation for sensible heat flux to deduce:
\begin{align}
    L_X = \left(\frac{\sqrt{Pr_{t0} Pr_t}}{c_\theta} \right)\left(\frac{\sigma_\theta}{\Gamma} \right),
    \label{LX}
\end{align}
where $L_X$ is the characteristic length scale for stably stratified flows. The standard deviation of potential temperature is $\sigma_\theta$. The turbulent Prandtl number for non-buoyant flows is denoted by $Pr_{t0}$; it is typically assumed to be equal to 0.85. The coefficient $c_\theta$ is approximately equal to 2.

In \cite{basu21c}, the dissipation rates of TKE ($\overline{e}$) and variance of temperature ($\sigma_\theta^2$) are derived to be respectively as follows:
\begin{subequations}
\begin{equation}
    \overline{\varepsilon} = \left(\frac{\sigma_w^3}{c_w^3 L_X}\right),
    \label{EDR5}
\end{equation}
and
\begin{equation}
    \overline{\chi}_\theta = \left(\frac{2 Pr_{t0}}{c_w c_\theta^2} \right) \left(\frac{\sigma_w \sigma_\theta^2}{L_X}\right).
    \label{CHI3}
\end{equation}
\end{subequations}
Here the standard deviation of vertical velocity is $\sigma_w$. The coefficient $c_w$ is approximately equal to 1.25. From Eqs.~(\ref{CT2a}), (\ref{EDR5}), and (\ref{CHI3}), we get: 
\begin{widetext}
\begin{equation}
    C_T^2 = \left(\frac{c}{2} \right) \left( \frac{c_w L_X^{1/3}}{\sigma_w} \right) \left(\frac{2 Pr_{t0}}{c_w c_\theta^2} \right) \left(\frac{\sigma_w \sigma_\theta^2}{L_X}\right) = \left(\frac{c Pr_{t0}}{c_\theta^2} \right)\left(\frac{\sigma_\theta^2}{L_X^{2/3}} \right).
    \label{CT2c}
\end{equation}
\end{widetext}
By substituting $\sigma_\theta$ from Eq.~(\ref{LX}) to Eq.~(\ref{CT2c}), we find: 
\begin{equation}
    C_T^2 = \left(\frac{c Pr_{t0}}{c_\theta^2} \right) L_X^{4/3} \Gamma^2 = \left(\frac{c}{Pr_t}\right) L_X^{4/3} \Gamma^2. 
    \label{CT2d}
\end{equation}
This equation is identical to Eq.~(\ref{CT2b}) with one important difference. Here the length scale $L_X$ is defined by Eq.~(\ref{LX}), and is not a free parameter. Several well-known length scales are explicitly related to $L_X$; please refer to Appendix~1 and \cite{basu21c} for further details.

Earlier, we mentioned that in the atmospheric optics literature the value of $Pr_t$ is frequently assumed to be equal to unity. However, there is ample evidence in the fluid dynamics and atmospheric turbulence literature that $Pr_t$ monotonically increases with increasing values of $Ri_g$. In this context, a physically-based formulation was derived by \cite{basu21c}. 

\section{$C_T^2$ in Convective Mixed Layer}
\label{CBL}

Convective (unstable) condition typically occur over land during daytime conditions. At offshore locations, such conditions prevail when air temperature is colder than the underlying sea-surface temperature. Under convective conditions, due to intense turbulent diffusion, the so-called 'mixed' layer (ML) develops a few tens of meters above the land or sea-surface. Due to turbulent mixing, (potential) temperature, moisture, and other meteorological variables become uniformly distributed in the ML. In other words,  in the ML, $\Gamma  = \frac{\partial \overline{\theta}}{\partial z} \approx 0$. As a result, Tatarskii's equation [i.e., Eq.~(\ref{CT2b})] and our proposed formulation [i.e., Eq.~(\ref{CT2d})] incorrectly predict $C_T^2 \approx 0$ in the ML. The reasons behind this unphysical prediction are discussed below.

In the ML, large coherent structures (called thermals, plumes) are primarily responsible for turbulent mixing. These structures originate near the surface and can reach the top of the ML (on the order of a km or so) in a matter of 10 to 15 minutes. During their ascent, these structures cause `non-local' mixing. As a result, turbulent fluxes are no longer proportional to the local mean gradients. In other words, the application of K-theory is not tenable in the ML. Hence, Eqs.~(\ref{uw}) and (\ref{wT}) are not valid in the ML \cite{holtslag91}. Furthermore, the simplified variance and flux budget equations [e.g., Eq.~(\ref{EDR4})] are not suitable for the ML. Additional non-local transport terms should be included in them. Unfortunately, such modifications make these equations analytically intractable. 

We recommend the readers to use Eq.~(\ref{CT2b}) and/or Eq.~(\ref{CT2d}) only for stably stratified conditions. Such conditions are omni-present in the free atmosphere (i.e., the layer above the atmospheric boundary layer). Also, nocturnal boundary layers over land are commonly stably stratified. For convective ML, a similarity-based formulation by Kaimal et al.~\cite{kaimal76} or its variants could be used. On this topic, an interesting study was recently published by \cite{luce20}. However, more research is highly desired in this arena. 

\section{Concluding Remarks}
\label{Conc}

Fifty years ago, Tatarskii developed a simple $C_T^2$ formulation which has found wide usage in a range of scientific and engineering disciplines, from astronomy to free-space optical communication. In this paper, we revisit this formulation and point out its limitations. We then propose a revised $C_T^2$ formulation based on turbulence variance and flux budget equations. In contrast to Tatarskii's equation, our formulation includes a novel length scale which is physically-based. Since this length scale is simply dependent on certain variances and mean gradients, it can be estimated from observational data. Such applications will be reported in a separate publication.

\begin{acknowledgements}
We are truly grateful to Hubert Luce and Lakshmi Kantha for thought-provoking communications and for providing constructive feedback on our work. 
\end{acknowledgements}

\section*{Appendix 1: Inter-relationships of Length Scales}
\label{Appendix}

In addition to Eq.~(\ref{LX}), two alternative definitions of $L_X$ were proposed by \cite{basu21c}:
\begin{subequations}
\begin{align}
    L_X & = c_H \left( \frac{\sigma_w}{S} \right) \left( \frac{1}{\sqrt{1/G - Ri_g/Pr_t}} \right), \label{LX1}\\
        & = c_H \left( \frac{\sigma_w}{N_{BV}} \right) \left( \frac{\sqrt{Ri_g}}{\sqrt{1/G - Ri_g/Pr_t}} \right).
    \label{LX2}
\end{align}
\end{subequations}
Where the coefficient $c_H$ is approximately equal to 0.8. The term $G$, called a `growth factor' \cite{schumann95,basu21c}, represents the ratio of the production and dissipation terms of the TKE equation. For weakly/moderately stable condition, $G$ equals to one. Under strongly stratified condition $G$ is expected to be smaller than one. Basu and Holtslag~\cite{basu21c} proposed the following heuristic parameterization for $G$:
\begin{equation}
    G = \min \left(1, Ri_g^{-1} \right).
\end{equation}

An unique relationship between $L_X$ and the Corrsin's length scale ($L_C$) can be established by utilizing Eq.~(\ref{LC}), Eq.~(\ref{EDR5}) and Eq.~(\ref{LX1}):
\begin{align}
    \frac{L_X}{L_C} & = \left( \frac{1}{\sqrt{1/G - Ri_g/Pr_t}} \right)^{3/2}. 
\end{align}
When $Ri_g \to 0$, $G = 1$ and $L_X \to L_C$. 

A different relationship can be deduced between the Ozmidov length scale ($L_{OZ}$) and $L_X$ via Eq.~(\ref{LOZ}), Eq.~(\ref{EDR5}) and Eq.~(\ref{LX2}):
\begin{align}
    \frac{L_X}{L_{OZ}} & = \left( \frac{\sqrt{Ri_g}}{\sqrt{1/G - Ri_g/Pr_t}} \right)^{3/2}. 
\end{align}
For $Ri_g \gg 1$, if $G$ is proportional to $Ri_g^{-1}$, then $L_X \to L_{OZ}$. 

In summary, the length scale $L_X$ acts as a smooth interpolator between two limiting length scales: $L_C$ and $L_{OZ}$. For this reason, it has been called a hybrid length scale. 

\bibliographystyle{spbasic_updated}      
\bibliography{Tatarskii}   

\end{document}